\documentclass[twocolumn,showpacs,floatfix,aps]{revtex4}

\usepackage{graphicx}
\usepackage{dcolumn}
\usepackage{hyperref}

\begin{document}

\title{
Neutron matter at zero temperature with auxiliary field 
diffusion Monte Carlo
}

\author{A. Sarsa}
\email{sarsa@sissa.it}
\altaffiliation{Present address: 
Departamento de F\'{\i}sica Moderna, Universidad de Granada, E-18071 Granada,
Spain}
\affiliation{International School for Advanced Studies, SISSA, 
and \\ INFM {\sl DEMOCRITOS} National Simulation Center \\ Via Beirut
I-34014 Trieste, Italy}
\author{S. Fantoni}
\email{fantoni@sissa.it}
\affiliation{International School for Advanced Studies, SISSA, 
and \\ INFM {\sl DEMOCRITOS} National Simulation Center \\ Via Beirut
I-34014 Trieste, Italy}
\author{K. E. Schmidt}
\email{kevin.schmidt@asu.edu}
\altaffiliation{Permanent address: Department of Physics and Astronomy, 
Arizona State University, Tempe, AZ, 85287}
\affiliation{International School for Advanced Studies, SISSA, 
and \\ INFM {\sl DEMOCRITOS} National Simulation Center \\ Via Beirut
I-34014 Trieste, Italy}
\author{F. Pederiva}
\email{pederiva@science.unitn.it}
\affiliation{Dipartimento di Fisica dell'Universit\`{a} di Trento, 
and \\ INFM {\sl DEMOCRITOS} National Simulation Center \\ 
I-38050 Povo, Trento, Italy}

\date{\today}

\begin{abstract}
The recently developed auxiliary field diffusion Monte Carlo
method is applied to compute the equation of state and the compressibility
of neutron matter.
By combining diffusion Monte Carlo for the spatial degrees of 
freedom and auxiliary field Monte Carlo to separate the spin-isospin
operators, quantum Monte Carlo
can be used to simulate the ground state of many nucleon systems
$(A\alt 100)$.
We use a path constraint to control the fermion sign problem.
We have made simulations for realistic interactions, which include tensor
and spin--orbit two--body potentials as well as three-nucleon forces. The
Argonne $v_8'$ and $v_6'$ two nucleon potentials 
plus the Urbana or Illinois three-nucleon potentials have been used in our 
calculations.  We compare with fermion hypernetted chain
results.
We report results of a Periodic Box--FHNC calculation, which
is also used to estimate the finite size corrections to our quantum
Monte Carlo simulations. 
Our AFDMC results for $v_6$ models of pure neutron matter 
are in reasonably good agreement with equivalent Correlated Basis Function
(CBF) calculations, providing energies per particle which are slightly lower
than the CBF ones. 
However, the inclusion of the spin--orbit force leads to quite different results
particularly at relatively high densities. The resulting equation of state
from AFDMC calculations is harder
than the one from previous Fermi hypernetted chain studies
commonly used to determine the neutron star structure.
     \end{abstract}

\pacs{26.60.+c,21.65.+f,21.30.Fe,05.10.Ln}
%26.60.+c Nuclear matter aspects of neutron stars
%21.65.+f Nuclear matter ... 
%21.30.Fe Forces in hadronic systems and effective interactions
%05.10.Ln Monte Carlo methods
\maketitle
\section{Introduction}

The important role played by N--N correlations on several properties of
dense and cold hadronic matter is 
a well established fact \cite{pandharipande97}. 
Less established are quantitative studies performed
with realistic nuclear interactions derived from N--N data and the spectra of 
light nuclei. The strong repulsion at short range accompanied with 
the strong spin--isospin dependence, make
\textit{ab initio} calculations of the nuclear matter  
equation of state  one of the 
most challenging problems in strongly correlated many--body theory.

A theoretical calculation of the nuclear matter energy per particle, 
as a function of the number density $\rho$,
the temperature $T$ and the neutron--proton 
asymmetry $\alpha = (N-Z)/(N+Z)$, with an 
uncertainty of less than an MeV has become a
fundamental issue. 
On one hand, one would like to use the 
observational data from neutron stars and supernovae, as well as
from heavy--ion collisions to get information on the many--body nature of the
nucleon interaction. On the other hand, it is of interest to understand
the effect of N--N correlations, and particularly of those induced by 
the tensor force, on the structure and the evolution of
compact astrophysical objects 
\cite{raffelt96,sawyer75,sawyer89,iwamoto82,reddy99}. 

In this paper we limit ourselves to
non--relativistic model hamiltonians.
Modern two--body potentials \cite{wiringa95,stoks94,machleidt96} 
fit the Nijmegen N--N data \cite{stoks93} below 350 MeV
at a confidence level of 
$\chi^2/N_{data}\sim 1$, and to a large extent give equivalent results
for several nuclear and neutron matter properties \cite{engvik97}. 
However, it has become evident that a two--body potential
alone is not sufficient to reproduce the experimental data of nuclei other 
than the deuteron ($A=2$). In the last few years, 
the Urbana--Argonne collaboration has produced three--body force
models which, when added to the two--body potential, provide a satisfactory
fit to the binding energies and the low--lying states of light nuclei
with $A\leq 10$ \cite{pudliner95,pieper01,pieper02}.

It would be desirable to have microscopic calculations of the equation 
of state of nuclear matter with an accuracy comparable to that of
light nuclei or, at least, on the order of the experimental uncertainties
of the equilibrium density, $\rho_0$, binding energy per particle 
at $\rho_0$ and  compressibility. This can be considered as the minimal 
requirement to attempt 
the study hadronic matter at densities larger than $\rho_0$, and/or
with large asymmetries ($\alpha$ close to $1$) in a realistic way.
Such calculations have to deal necessarily with potentials
which are strongly spin--isospin dependent and which include a three--body
force. 

Most of the microscopic calculations of the nuclear matter equation of
state
carried out in the last decades have been performed
by using perturbation theories based either on  
ladder diagram summation, like
Brueckner or Green's Function theories \cite{engvik97,ramos91},
or Correlated Basis Function
theories, based on Fermi hypernetted chain techniques 
\cite{wiringa88,akpa97,morales02}. In spite of
the important advances made in recent years in the above theories, 
the required accuracy for the equation of state has not yet been reached.

Quantum Monte Carlo methods have been very successful in calculating the
properties of strongly interacting systems in condensed matter physics. They 
are substantially exact, apart from statistical errors, finite 
size effects and the well known 
sign problem \cite{schmidt84} for Fermi systems. They have been recently 
used to perform quantum simulations of light
nuclei \cite{pudliner97,pieper98,pieper02} with modern nonrelativistic
Hamiltonians of the type discussed above.  However, the exponential
growth in the number of spin-isospin states with the number of nucleons
$A$, has kept this
method from being applied to larger nuclear systems.

Auxiliary field diffusion Monte Carlo \cite{schmidt99} (AFDMC) has
been especially developed to tackle 
the problem of computing the binding energy of a relatively large nuclear 
system at the required accuracy.
In this approach the particle coordinates are propagated as in
standard diffusion Monte Carlo. Auxiliary fields are introduced to
uncouple the spin-dependent interaction between particles by means of a 
Hubbard--Stratonovich transformation. The particle spins only interact with
the auxiliary fields which, when integrated, produce the original interaction.
The method consists of calculating the auxiliary field integrations by
Monte Carlo sampling and then propagating the spin
variables. This propagation results in a rotation of each particle's 
spinor governed by the sampled values of the auxiliary variables.
The result is a sampling of the spin variables which should have
less variance than a direct approach where the spins are flipped.

The tensor force couples the spin configurations with
the orbital angular momentum so that the wave function becomes complex.
The resulting fermion phase problem is handled by applying a path-constraint
approximation analogous to the fixed-node approximation. 
The AFDMC method for the spin--isospin calculations
can be viewed as a generalization of the method of Zhang et al. 
\cite{zhang95,zhang97}
used in condensed matter lattice systems to the spin-isospin states of nucleon
systems, while retaining standard diffusion Monte Carlo for the spatial
degrees of freedom. The AFDMC method has proved to be efficient in dealing
with large nucleon systems interacting via semi-realistic potentials
\cite{schmidt99,manchester,fantoni00} and spin-polarized neutron 
systems \cite{fantoni01b}.

The aim of this paper is to give a detailed description of the AFDMC method
and to report results for the equation of state
of pure neutron matter ($\alpha = 1$) with a fully realistic nuclear
interaction, at zero temperature. 
It presents results of AFDMC simulations of 14, 38, 66 and
114 neutrons in a periodic box, interacting via a realistic potential
which includes two--body tensor and spin--orbit components, as well as
three--body forces. Particular attention is paid to the 14
neutron system, which may serve as a homework problem for 
different many--body techniques. It is small enough to be 
handled by traditional quantum Monte Carlo methods \cite{carlson03}.
However, it will be shown that the finite size effects of 14 neutron systems
are hard to estimate in a realistic way. Actually results obtained with 
larger systems (66 or 114 neutrons) show that the equation of state of neutron
matter cannot be simulated starting from 14 neutron in a box, particularly
in the high density region.
Finite size effects for the larger systems considered here can be fairly
well estimated by the recently developed Periodic Box FHNC (PBFHNC) 
theory \cite{fantoni01}.
We have also performed AFDMC calculations of the binding energy of
symmetric and asymmetric nuclear matter. A few results obtained with 
semi-realistic spin-dependent central potentials are presented and
discussed.

The plan of the paper is the following. The Hamiltonian used in this work
is shown in the next section.  In section \ref{sec.spin} the problem of
the spin degrees of freedom in quantum monte carlo simulations is discussed.
Section \ref{sec.afdmc} is devoted to the
description of the AFDMC method, including the calculation of the spin--orbit
and the three--body terms of the Hamiltonian. A discussion of the 
finite size effects along with the Periodic box FHNC method is given
in Section \ref{sec.fhnc}. 
The results for the neutron matter equation of state
are presented and
discussed in Section \ref{sec.results}.
The conclusions and perspectives for the present
work are in Section \ref{sec.conclusions}.

\section{The Hamiltonian}
\label{sec.hamiltonian}

We use a non relativistic Hamiltonian of the form

\begin{eqnarray}
H & = & T+V_2+V_3 \nonumber \\
& = &-\frac{\hbar^2}{2m}\sum_{j=1,N}\nabla^2_j + 
\sum_{j<k}v_{jk} + \sum_{j<k<l} V_{jkl} \ ,
\label{ham_ham}
\end{eqnarray}
containing the kinetic term, where we have used 
$\hbar^2/(2m)=20.73554$  MeV fm$^2$ (which corresponds to the $n-p$ 
reduced mass), and  
two-- and three--body potentials.  The two--body potential belongs to the
Urbana--Argonne $v_\ell$ type

\begin{eqnarray}
v_\ell = \sum_{j<k}v_{jk} =
\sum_{j<k}\sum_{p=1}^\ell v_p(r_{jk}) O^{(p)}(j,k)\ ,
\label{ham_arg}
\end{eqnarray}
where $j$ and $k$ label the two nucleons, $r_{jk}$ is the distance
separating the two nucleons, and the spin--isospin dependent operators 
$O^{p}(i,j)$ for $p=1,8$ are given by

\begin{eqnarray}
O^{p=1,8}(j,k)  =  \left(1, {\vec \sigma}_j\cdot{\vec \sigma}_k, S_{jk},
 {\vec L}_{jk}\cdot{\vec S}_{jk} \right)
\otimes \left(1,{\vec \tau}_j\cdot{\vec \tau}_k\right) \ , 
\label{ham_ope}
\end{eqnarray}
where $S_{jk} = 3 (\hat{r}_{jk}\cdot{\vec \sigma}_j)
(\hat{r}_{jk}\cdot{\vec \sigma}_k)
-{\vec \sigma}_j\cdot {\vec \sigma}_k$ is the two--nucleon tensor operator,
and ${\vec L}_{jk}$ and ${\vec S}_{jk}$ are the relative angular momentum and 
the total spin, given by

\begin{eqnarray}
{\vec L}_{jk} & = & \frac{\hbar}{2\imath} ({\vec r}_j - {\vec r}_k) \times
({\vec \nabla}_j - {\vec \nabla}_k)\ , \\
{\vec S}_{jk} & = & 
\frac{\hbar}{2} ( {\vec \sigma}_j + {\vec \sigma}_k) \ .
\label{ham_so}
\end{eqnarray}

The full Argonne $v_{18}$ 
potential consists of $\ell$=18 components.
Besides the 8 components given in  
Eq.(\ref{ham_ope}), it includes the 6  $(L^2$,
$L^2\ {\vec \sigma}_j\cdot {\vec \sigma}_k$, $({\vec L}\cdot{\vec S})^2)$
$\otimes (1,{\vec \tau}_j\cdot{\vec \tau}_k)$ charge independent ones, 
as well as 4 other charge--symmetry--breaking and charge--dependent terms.

We use a simplified isoscalar version of the $v_{18}$ potential, 
the so called $v_8'$ two--body potential \cite{pudliner97}. 
This potential has been obtained with a new fit to the N--N data,
with only the first eight spin--dependent operators in
Eq.(\ref{ham_ope}) included. It equals the isoscalar part of $v_{18}$ 
in all $S$ and $P$ waves as well as in the $^3D_1$ wave and
its coupling to the $^3S_1$. It has been used in a number of GFMC calculations 
in light nuclei \cite{pudliner97}, 
as well as FHNC/SOC calculations in nuclear matter \cite{akpa97};
differences with the
$v_{18}$ potential give small contributions and 
can be safely estimated perturbatively or from
FHNC/SOC calculations.

For the sake of completeness, we report here the parameterization 
of the Argonne $v_8'$ two--body potential.

\begin{eqnarray}
v_p(r) = \sum_{m=1}^8 A_{p,m} F_m(r) \ ,
\label{ham_pot}
\end{eqnarray}

where odd and even components refer to ${\tau}$--independent and
${\tau}$--dependent operators respectively. The spin--independent part
$v_{ij}^{SI}$ of the two--body potential 
is given by the first component $v_{p=1}(r_{ij})$ only.
The constants $A_{p,m}$ and the functions $F_m(r)$ are given in the 
appendix.

In the case of pure neutron matter (PNM), 
the isospin exchange operators are replaced
by the identity. 

We denote by $v_6'$ the two--body potential model obtained
by restricting the $v_8'$ potential to the first 6 (3 for neutron matter)
components.
Note that this truncation of the Argonne $v_8'$ should not be confused
with the recently produced Argonne $AV6'$ potential \cite{wipi03}.

The three-body interaction used in our calculations of the equation of state
is the Urbana IX potential \cite{pudliner97}. 
For neutrons, the Urbana-IX
interaction is given by the sum of a spin independent and a
spin dependent part

\begin{eqnarray}
V_{jkl} = V_{jkl}^{SI} + V_{jkl}^{SD} \ ,
\label{v3}
\end{eqnarray}

where

\begin{eqnarray} 
 V_{jkl}^{SI} &=& U_0 \sum_{cyclic} 
 T^2(m_{\pi},c_3;r_{jl}) T^2(m_{\pi},c_3;r_{lk}) \ , \nonumber \\
 V_{jkl}^{SD} &=& B_{2\pi} \sum_{cyclic} 
\{ X_{jl}^{\pi},X_{lk}^{\pi}\}  \ ,
\label{ham_three}
\end{eqnarray}

and the operator $X_{jk}^{\pi}$ is given by

\begin{eqnarray}
X_{jk}^{\pi} = Y(m_{\pi},c_3;r_{jk}) {\vec \sigma}_j\cdot {\vec \sigma}_k
+ T(m_{\pi},c_3;r_{jk}) S_{jk}\ .
\end{eqnarray}

The values of the parameters of the Urbana IX three--body potential, used in
our calculations are shown in the appendix. Notice that 
in some of our earlier AFDMC calculations we have used $c_3= 2.0$ fm$^{-2}$ 
and $\mu=0.7$ fm$^{-1}$, as 
given in the original papers proposing the Urbana IX 
potential \cite{carlson83} and the $v_8'$ model interaction \cite{pudliner95}.
Changing $c_3$ from $2.0$ to $2.1$ leads to a $\sim 10\%$ additional increase 
of the three-body force contribution in neutron matter.
In the following, 
we will denote with $AU8'$ the $v_8'$ plus Urbana IX interaction, with 
$AU6'$ the $v_6'$ plus Urbana IX interaction. 

We have also considered the recently developed Illinois three-body 
potentials, which 
include two $\Delta$ intermediate state diagrams \cite{pieper01}, and denoted
with IL1,$\ldots$, IL4.

\section{Spin Degrees of Freedom}
\label{sec.spin}

Standard Green's function or diffusion Monte Carlo methods for central
potentials sample only the particle positions since the spin or isospin
of the particles can be fixed. The Green's function Monte Carlo method
used in light nuclei also samples the particle positions, but a complete
description of the spin degrees of freedom is kept for each position
sample leading to an exponential growth of the number of spin-isospin
states with particle number $A$. This exponential behavior can be removed
by sampling rather than summing the spin-isospin degrees of freedom.

We define a walker to be the $3A$ coordinates of the $A$ particles and
$A$ spinors each giving the four amplitudes for a particle to be in the
proton up, proton down, neutron up and neutron down states. For the
special case where walkers are sampled from the usual neutron-proton
up-down basis, the spinors would be one of $(1,0,0,0)$, $(0,1,0,0)$,
$(0,0,1,0)$, and $(0,0,0,1)$ for each particle.
Our auxiliary field method requires the more general definition as
shown below.

As usual,
the overlap of the walker bra with the trial ket is the wave function
amplitude,
\begin{equation}
\langle R,S | \Psi_T \rangle \equiv \Psi_T(R,S) \,.
\end{equation}

Direct sampling of the spin-isospin in the usual spin up/down basis
requires a good trial function that can be evaluated efficiently.
This can be most easily seen for the variational formalism, but the
same analysis applies to Green's function or diffusion Monte Carlo.
A variational Monte Carlo calculation can be formulated by minimizing
the expectation value of the Hamiltonian,
\begin{eqnarray}
\langle H \rangle &=& \frac{\langle \Psi_T | H | \Psi_T \rangle}
{\langle \Psi_T | \Psi_T \rangle}
\nonumber\\
&=& \frac{\int dR \sum_{S,S'} \Psi_T^*(R,S') H_{S',S} \Psi_T(R,S)}
{\int dR \sum_S |\Psi_T(R,S)|^2} \, ,
\end{eqnarray}
where for a $v_6$ interaction we would have
\begin{equation}
H_{S',S} = \langle S'|S \rangle \left [ -\frac{\hbar^2}{2m} \sum_n \nabla^2_n
\right ] + \langle R S' | V | R S \rangle \, ,
\end{equation}
with a straightforward generalization for spin-orbit terms.

Variational Monte Carlo can be implemented with either spin
sums \cite{lomnitz81,carlson88,buendia00}
\begin{eqnarray}
\langle H \rangle &=& \int dR E_L(R) P(R) \, ,
\nonumber\\
P(R) &=& \frac{\sum_S |\Psi_T(R,S)|^2}{\int dR \sum_S |\Psi_T(R,S)|^2} \, ,
\nonumber\\
E_L(R) &=& \frac{\sum_{S,S'} \Psi_T^*(R,S')H_{S',S} \Psi_T(R,S)}
{\sum_{S} |\Psi_T(R,S)|^2} \, ,
\end{eqnarray}
or spin samples \cite{caka85}
\begin{eqnarray}
\langle H \rangle &=& \int dR \sum_S E_L(R,S) P(R,S) \, ,
\nonumber\\
P(R,S) &=& \frac{|\Psi_T(R,S)|^2}{\int dR \sum_S |\Psi_T(R,S)|^2} \, ,
\nonumber\\
E_L(R,S) &=& \frac{\sum_{S'} \Psi_T^*(R,S')H_{S',S} \Psi_T(R,S)}
{|\Psi_T(R,S)|^2} \, .
\end{eqnarray}
In these equations, $P$ is the probability density to be sampled
and $E_L$ is the local energy. A typical variational calculation
would use the Metropolis algorithm to sample either $R$ or $R$ and $S$ from $P$,
and average the value of the local energy over these samples.

Notice that for an eigenstate of $H$, both $E_L(R,S)$ and $E_L(R)$ are
constant. So, as for central potentials, the variance will be low if
the trial function is accurate. Notice also that the spin sum $S'$ in
the definition of $E_L(R,S)$ is polynomial rather than exponential
in $A$. For example a pair
potential will have only order $A^2$ terms where two particles have different
spin-isospin.

The variance per sample for complete spin sums
will be lower than for spin samples. However, since the spin sums grow
exponentially with particle number spin sampling should be more efficient
for large particle number if the trial function can be evaluated efficiently
for a single many-particle spin state $S$.

Unfortunately, all of the good trial wave functions currently used
for large numbers of
particles cannot be evaluated efficiently for a single many-particle
spin state $S$.
For example light nuclei variational Monte Carlo calculations
are typically done using a pair product (or more complicated)
wave function,
\begin{equation}
| \Psi_P\rangle = {\cal S} \prod_{j<k} f^c_{jk}
\left [1 + \sum_p u^p_{jk} O^p_{jk} \right ]
|\Phi \rangle \, ,
\end{equation}
where ${\cal S}$ symmetrizes the operator products, and $|\Phi\rangle$
is the antisymmetric model state. While the symmetrizer produces all
possible orderings of the operators and therefore gives ${\cal O} (A^2!)$
terms, normally the commutator terms are fairly small and the ordering
of the operators is sampled. However, even within a fixed ordering, each
operator in the product term when operating on
a single many-particle spin-isospin state will produce 4 or 8 new states
depending on whether isospin exchange gives a new state. ${\cal O} (A)$ 
operators out of the ${\cal O}(A^2)$ total acting on a single state 
are enough to populate all the states. Therefore a
straightforward evaluation of $\langle R S| \Psi_P\rangle$, for this
wave function will have the same computational complexity as evaluating
a complete set of spin-isospin states at the position $R$.
Since computing all the states have the same cost as a single state,
full spin sums are used for these calculations.

If good trial functions for spin-isospin dependent interactions can
be devised which can be evaluated or sampled
efficiently at a single many-particle
space position and spin-isospin state, straightforward
generalizations of standard central potential Monte Carlo methods,
both variational and Green's function, with spin-state sampling
will solve the nuclear many-particle
Hamiltonian.

\section{The AFDMC method}
\label{sec.afdmc}

Since direct evaluation of the pair product wave function is not
computationally feasible for large numbers of particles, and so far we 
have no good methods of sampling these wave functions, we instead
drop the operator terms
altogether and sample the spin-isospin variables
using a rather poor, but easy to
evaluate, wave function. Since this wave function does not contain
amplitudes of the spin states of the correct solution we cannot use it
to sample the spins. Instead, we rewrite the propagator as an
integral over auxiliary
fields using the
Hubbard-Stratonovich transformation
\begin{eqnarray}
\label{eq.hs}
e^{-\frac{1}{2} \lambda O^2 \Delta t}
= \frac{1}{\sqrt{2\pi}} \int_{-\infty}^\infty dx e^{-\frac{x^2}{2}}
e^{x\sqrt{- \lambda \Delta t}O} \, ,
\end{eqnarray}
where $O$ can be a one-body operator. To make use of this transformation we
write our propagator as the left-hand side of Eq. \ref{eq.hs}, so that
the integrand of the right hand side is a product of one-body terms.
The integrand has a form such that propagating a walker at
$|R,S\rangle$ results in another walker of the same form at $|R',S'\rangle$.

For $N$ neutrons, the $v_6$ two-body interaction can be split into
two parts
\begin{eqnarray}
\sum_{j<k} v_{jk} = B+\frac{1}{2}\sum_{j,\alpha,k,\beta} \sigma_{j,\alpha}~
A_{j,\alpha;k,\beta} ~\sigma_{k,\beta} \, ,
\end{eqnarray}
where roman subscripts like $j$ and $k$ are particle labels while greek
subscripts like $\alpha$ and $\beta$ are cartesian components.
The matrix $A$ and the scalar $B$ are functions of the particle positions,
\begin{eqnarray}
B &=& \sum_{j<k} \left [ v_1(r_{jk})+v_2(r_{jk}) \right ] \, ,
\nonumber\\
A_{j,\alpha;k,\beta} &=& 
(v_3(r_{jk})+v_4(r_{jk}))\delta_{\alpha\beta} +
\nonumber\\
&&
\left [ v_5(r_{jk})+v_6(r_{jk}) \right ]
\left [3 \hat{r}_{jk} \cdot \hat x_{\alpha}\,
         \hat{r}_{jk} \cdot \hat x_{\beta} 
        -\delta_{\alpha\beta} \right ].
\nonumber\\
\label{matrixA}
\end{eqnarray}
$A_{j,\alpha;k,\beta}$
is taken to be zero when $j=k$. $A$ can be viewed as a $3N$ by $3N$
real symmetric matrix.  
It therefore has real eigenvalues and eigenvectors defined by
\begin{eqnarray}
\sum_{k,\beta} A_{j,\alpha;k,\beta } \psi_n^{k\beta} = \lambda_n
\psi_n^{j\alpha}  \,.
\end{eqnarray}
The potential can be written as
\begin{eqnarray}
\label{eq6}
\sum_{j<k} v_{jk} &=& B+\frac{1}{2}\sum_{j,\alpha,k,\beta,n}
\sigma_{j\alpha} \psi_n^{j\alpha}
\lambda_n \sigma_{k\beta} \psi_n^{k\beta}
\nonumber\\
&=& B+ \frac{1}{2} \sum_{n=1}^{3A} ( O_n )^2 \lambda_n \ ,
\end{eqnarray}
with
\begin{eqnarray}
O_n = \sum_{j\alpha} \sigma_{j\alpha} \psi_n^{j\alpha} \ .
\end{eqnarray}
Each of the $O_n$ is a sum of 1-body operators as required above.

After applying the Hubbard-Stratonovich 
transformation, the short time approximation for the
propagator can be written as

\begin{eqnarray}
&&\left ( \frac{m}{2 \pi \hbar^2\Delta t} \right )^{3A/2}
\exp\left ( -\frac{m |R - R'|^2}{2 \hbar^2\Delta t} \right )
e^{-B(R) \Delta \tau}
\nonumber\\
&&\prod_n \frac{1}{\sqrt{2\pi}} \int_{-\infty}^\infty dx_n 
e^{-\frac{x_n^2}{2}}
e^{x_n\sqrt{- \lambda_n \Delta t}O_n} \, .
\end{eqnarray}

The $O_n$ do not commute, so we need to keep the time steps small so that
the commutator terms can be ignored.

We sample a value of  $x$ for each of the $3A$ auxiliary field variables.
Once these values are known, the propagation reduces to  a
rotation in the spin space, and, therefore, to multiplying
the current spinor value for each particle by the set of 
matrices given by the transformation above. For a given eigenvalue 
$\lambda_n \leq 0$ 
in Eq.(\ref{eq6}) the spin  states of particle $k$, 
$| \eta_k' \rangle = a_k'|\uparrow \rangle + b_k'|\downarrow \rangle$  
will be rotated to the
new one $| \eta_k \rangle $ having the following components

\begin{eqnarray}
a_k &=& a_k'(\cosh(\alpha_n) + \sinh(\alpha_n)\psi_n^z(k))\nonumber \\
 &+& b_k' \sinh(\alpha_n)
  (\psi_n^x(k) - \imath \ \psi_n^y(k)) , \nonumber \\
b_k &=& b_k'(\cosh(\alpha_n) -\sinh(\alpha_n)\psi_n^z(k))\nonumber \\
 &+& a_k' \sinh(\alpha_n)
  (\psi_n^x(k) +\imath \ \psi_n^y(k)) ,  
\label{rotation}
\end{eqnarray}
where
\begin{eqnarray}
\alpha_n = \Delta t |\lambda_n| x_n 
\sqrt{(\psi_n^x(k))^2+(\psi_n^y(k))^2+(\psi_n^z(k))^2} \ ,
   \end{eqnarray}

and $x_n$ is the sampled Hubbard--Stratonovich value.
For positive values of $\lambda_n$, one has a similar set of equations, in
which  $\sinh(\alpha_n)$ is substituted with $\imath \sin(-\alpha_n)$.

Finally it is worth mentioning here that importance sampling has been used for
the integral in the HS variables. The value of the overlap of the walker
with the trial function will not be peaked around $x_n=0$, but will be
shifted. Rather than sampling from the gaussian we preferentially sample 
values where we predict the trial function will be larger. One way is
to shift the sampled gaussian values with a drift term analogous to the
drift term in diffusion Monte Carlo by replacing the $\sigma$ operators
by their expectation value at the current $R,S$ value
and taking the real part. That is we write
\begin{eqnarray}
&&\frac{1}{\sqrt{2\pi}} \int_{-\infty}^\infty dx_n 
e^{-\frac{x_n^2}{2}}
e^{x_n\sqrt{- \lambda_n \Delta t}O_n} \, ,
\nonumber\\
&=&\frac{1}{\sqrt{2\pi}} \int_{-\infty}^\infty dx_n 
e^{-\frac{(x_n-\bar x_n)^2}{2}}
e^{x_n\sqrt{- \lambda_n \Delta t}O_n} e^{\frac{-2\bar x_n x_n +\bar x_n^2}{2}}
\, ,
\nonumber\\
&&\bar x_n = {\rm Re} \left [\sqrt{- \lambda_n \Delta t}~\langle O_n\rangle
\right ] \, ,
\nonumber\\
&&\langle O_n \rangle = \frac{\langle \Psi_T | O_n |R,S \rangle}
{\langle \Psi_T |R,S\rangle} \, ,
\label{drift}
\end{eqnarray}
and sample the shifted gaussian, the last correction term is included in
the weight. With this real shift and the compensating weight, only the
efficiency of the algorithm is changed. We have tried other schemes using
a discretized gaussian integration with altered probabilities
and compensating weights with very little difference in the overall
efficiency. In Ref. \cite{zhang03} a complex drift
rather than the real drift  in Eq. (\ref{drift}) has been
used. Unlike our real drift above, this
can change how the path constraint is applied. 

\subsection{Three--body potential}

For a neutron system the spin--dependent part of Urbana IX potential, 
given in eqs. (\ref{v3}) and (\ref{ham_three}) reduces to a 
sum of terms containing only two-body spin operators but
with a form and strength that depends on the positions of three
particles. As will be seen below,
for a fixed position of the particles, 
the inclusion of three-body
potentials of the Urbana IX type in the Hamiltonian 
does not add any additional complications. It simply changes
the strength of the coefficients of the terms in the potential
and  can be trivially incorporated in the AFDMC calculations.

The anticommutator in Eq.(\ref{ham_three}) can be written as

\begin{eqnarray}
\{X_{jl}^{\pi},X_{lk}^{\pi}\} = 2\ x_{jkl}^{\mu\nu}  
\sigma_j^{\mu} \sigma_k^{\nu} \ , 
\end{eqnarray}
where
\begin{eqnarray}
x_{jkl}^{\mu\nu} =   y_{jl}y_{lk}\delta_{\mu\nu}
+ y_{jl}t_{lk}^{\mu\nu} + t_{jl}^{\mu\nu}y_{lk} +
 t_{jl}^{\mu\alpha} t_{lk}^{\alpha\nu} \ , 
\end{eqnarray}
and
\begin{eqnarray}
y_{jl} &=& Y(m_{\pi},c_3,r_{jl})-T(m_{\pi},c_3,r_{jl}), \nonumber \\
t_{jl}^{\mu\nu} &=& 3\ T(m_{\pi},c_3,r_{jl}) \hat {r}_{jl}^{\mu}
\hat {r}_{jl}^{\nu}\ .
\end{eqnarray}

The spin--dependent part of the three--body interaction $V_3^{SD}$ can
then be easily incorporated in the matrix $A_{j,\alpha,k,\beta }$ of 
Eq.(\ref{matrixA}), by the following substitution

\begin{eqnarray}
A_{j,\alpha;k,\beta } \rightarrow A_{j,\alpha;k,\beta } + 
2\ \sum_l B_{2\pi} x_{jkl}^{\alpha\beta} \ .
\end{eqnarray}

Similarly the new terms in the Illinois potentials can be included into this
matrix.

\subsection{The Spin-Orbit Propagator}

A first order approximation \cite{pieper98} to the spin-orbit contribution to
the propagator can be obtained 
by operating the derivative appearing in the $\vec L_{jk} \cdot \vec S_{jk}$ 
operator on the free propagator $G_0$

\begin{eqnarray}
(\vec \nabla_j - \vec \nabla_k)&& \ G_0(R,R') = \nonumber \\
&& -\frac{m}{\hbar^2\Delta t} (\Delta \vec r_j - \Delta \vec r_k) G_0(R,R')\ ,
\end{eqnarray}

and substituting this expression back into the propagator. As a result, 
the spin--orbit part $P_{LS}$ of the propagator is factored out and is
finally written as

\begin{eqnarray}
P_{LS} &=& 
\exp\left (\sum_{j \ne k}\frac{m v_{LS}(r_{jk}) }{4 \imath \hbar^2}
[\vec r_{jk} \times (\Delta \vec r)_{jk}]
\cdot \vec \sigma_j \right )
\nonumber\\
&=& 
\exp\left (\sum_{j \ne k}\frac{m v_{LS}(r_{jk}) }{4 \imath \hbar^2}
( \vec \Sigma_{jk} \times
\vec r_{jk} ) \cdot \Delta \vec r_j \right ) 
\label{plsprop}
\end{eqnarray}
where $(\Delta \vec r)_{jk}=\Delta \vec r_j -\Delta \vec r_k$ and
$\vec \Sigma_{jk} = \vec \sigma_j + \vec \sigma_k$. 

However a careful analysis of the above expressions show that they include
some spurious contributions linear in $\Delta t$. In order to see this the
wave function is expanded, as usual, in the integral form of the
imaginary time Schr\"odinger equation
keeping only linear terms

\begin{eqnarray}
\label{eq8}
\Psi(R) &=& \Delta t \left [ \frac{1}{2m} \sum_j \nabla^2_j - V  + E_0
\right ] \Psi(R)
\nonumber\\
&+& \int dR' G_0(R,R') P_{LS} [ \Psi(R)  \nonumber \\
& -& \sum_p \Delta \vec r_p \cdot \vec \nabla_p \Psi(R)] + \ldots
\end{eqnarray}

At this point, $P_{LS}$ is expanded by using the second form of this propagator
given in Eq.(\ref{plsprop}) keeping both linear and quadratic terms in
$\Delta \vec r$. The integral in
$R^{\prime}$ can be done by taking into account that i) the gaussian integrates
to one if there are no powers of $\Delta \vec r $; ii) terms containing
only one power of a $\Delta \vec r$ integrate to zero; iii) 
quadratic terms containing powers of different components of $\Delta R'$,
integrate to zero and iv) terms like $(\Delta x_j')^2$ integrate 
to $\Delta t \hbar^2/m$. 

We first consider the part coming from the linear terms in $\Delta \vec r$ 
in both the wave function and $P_{LS}$. These terms, after integration give

\begin{eqnarray}
- \Delta t \sum_{j \ne k}\frac{v_{LS}(r_{jk}) }{4 \imath}
[( \vec \sigma_j +\vec \sigma_k) \times
\vec r_{jk} ] \cdot \vec \nabla_j \Psi(R) \, .
\nonumber \\
\end{eqnarray}

The expression above can be further simplified by 
interchanging the dummy indices  $j$ and $k$

\begin{eqnarray}
-\Delta t \sum_{j < k} v_{LS}(r_{jk}) [\vec L \cdot \vec S]_{jk}
\Psi(R) \ ,
\end{eqnarray}
which is the spin-orbit contribution to the
Hamiltonian multiplied by $-\Delta t$.

However the $P_{LS}$ propagator includes other terms which are of the
same order in $\Delta t$. They come from the quadratic $\Delta \vec r$ terms
of the expansion of $P_{LS}$

\begin{eqnarray}
&&\Delta t (V_2+V_3) = \Delta t \frac{m}{32}
\sum_j \sum_{k \ne j}\sum_{p \ne j}
v_{LS}(r_{jk}) v_{LS}(r_{jp}) \nonumber \\
&& ( \vec \Sigma_{jk} \times \vec r_{jk} )
\cdot
( \vec \Sigma_{jp} \times \vec r_{jp} )
\nonumber\\
&=&
\Delta t \frac{m}{32\hbar^2}
\sum_j \sum_{k \ne j}\sum_{p \ne j}
v_{LS}(r_{jk}) v_{LS}(r_{jp})
\nonumber\\
&&
\left \{
\vec r_{jk} \cdot \vec r_{jp}
\vec \Sigma_{jk} \cdot \vec \Sigma_{jp}
- \vec \Sigma_{jp} \cdot \vec r_{jk} \vec \Sigma_{jk}\cdot \vec r_{jp}
\right \} \ .
\nonumber\\
\end{eqnarray}

The terms with
$k=p$ give rise to a two--body additional effective potential $V_2^{add}=-V_2$,
 
\begin{eqnarray}
V^{add}_2 &=& -\sum_{j<k} 
\frac{m r_{jk}^2v^2_{LS}(r_{jk})}{8\hbar^2} [
2 + \vec \sigma_j \cdot \vec \sigma_k \nonumber \\
&-& \vec \sigma_j \cdot
\hat r_{jk} \vec \sigma_k \cdot \hat r_{jk} ] \ .
   \end{eqnarray} 

The terms with $k \ne p$ lead to a three--body additional effective potential
$V^{add}_3 = -V_3$, given by

\begin{eqnarray}
 V^{add}_3 &=& -\sum_{j<k<p} \sum_{\rm cyclic}
\frac{m r_{jk} r_{jp} v_{LS}(r_{jk}) v_{LS}(r_{jp})}{16\hbar^2}
\nonumber\\
&&
\{
\hat r_{jk} \cdot \hat r_{jp} \left [
2 + \vec \sigma_k \cdot \vec \sigma_j + \vec \sigma_p \cdot \vec \sigma_j
+\vec \sigma_k \cdot \vec \sigma_p \right ]
\nonumber\\
&-&\vec \sigma_j \cdot \hat r_{jk} \vec \sigma_k \cdot \hat r_{jp}
-\vec \sigma_p \cdot \hat r_{jk} \vec \sigma_j \cdot \hat r_{jp} \nonumber \\
&-&\vec \sigma_p \cdot \hat r_{jk} \vec \sigma_k \cdot \hat r_{jp}
\} \ .
\end{eqnarray}

Therefore in the actual propagation it is necessary to include explicitly
these terms with opposite sign if one is using $P_{LS}$ as given by
Eq.(\ref{plsprop}).

An alternative method that we have also used comes from realizing that
the counter terms are produced by the next order term in the series
expansion of the exponential. These terms either average to zero,
or are higher order in the time step or give incorrect contributions.
Subtracting them gives the propagator,
\begin{eqnarray}
&&\exp\left (\sum_{j \ne k}\frac{m v_{LS}(r_{jk}) }{4 \imath \hbar^2}
[\vec r_{jk} \times (\Delta \vec r)_{jk}]
\cdot \vec \sigma_j \right )
\nonumber\\
&&\exp\left (-\frac{1}{2}
\left [\sum_{j \ne k}\frac{m v_{LS}(r_{jk}) }{4 \imath \hbar^2}
[\vec r_{jk} \times (\Delta \vec r)_{jk}]
\cdot \vec \sigma_j \right ]^2 \right ) \, ,
\nonumber\\
\end{eqnarray}
with the second exponential giving the required counter terms to include.
The two forms are equivalent to first order in $\Delta t$.

\subsection{Trial wave function}

In our calculations we use the simple trial function 
given by a Slater determinant
of one-body space-spin orbitals multiplied by a central Jastrow
correlation,

\begin{eqnarray}
|\Psi_T \rangle =
\left [\prod_{j<k} f(r_{jk}) \right ]
{\cal A}\left [ \prod_j |\phi_j , s_j \rangle \right ] \,,
\label{jastrow}
\end{eqnarray}

where ${\cal A}$ is the antisymmetrizer of  $A$ particles.
The overlap of a walker with this wave function is
the determinant of the space-spin orbitals, evaluated
at the walker position and spinor for each particle (for nuclear matter the
spinor also includes the isospin), and multiplied
by a central Jastrow product.

For unpolarized neutron matter in a box of side $L$, the orbitals are plane
waves that fit in the box times up and down spinors.
The usual closed shells are 2, 14, 38, 54, 66, 114, $\ldots$ for neutrons and
4, 28, 76,  $\ldots$ for nucleons.

The Jastrow correlation function $f(r)$ has been taken as the first component
of the FHNC/SOC correlation operator $\hat F_{ij}$, which minimizes
the energy per particle of either neutron or nuclear matter at the desired 
density \cite{wiringa88} (see also Section \ref{sec.fhnc}). 

As noted in section \ref{sec.spin}, a trial function with spin exchange and
tensor correlations requires exponentially increasing computational
work as the number of particles increases. The advantages of our trial
function is that it is totally antisymmetric and for $A$ particles
requires order $A^3$ operations to evaluate. However, it does not
contain any amplitude generated by the tensor force where spins are
flipped with a compensating orbital angular momentum. It is left
to the AFDMC method to generate these missing components.

Other forms of a trial wave function can be used. For example including a linear
combination of Slater determinants is possible as is modifying the
orbitals to include spin correlations of backflow form \cite{fantoni02}.
Both of these avoid the exponential
computational complexity, but may not capture the essential physics
of the tensor force \cite{brualla03}. 

\subsection{Path Constraint}

As in standard fermion diffusion Monte Carlo, the AFDMC method has 
a fermion sign problem.
The overlap of our walkers with the trial function
will be complex in general so the usual fermion sign problem becomes
a phase problem.

To deal with this problem, we 
constrain the path of the walkers to regions where the real part of
the overlap with our trial function is positive. We have also tried
constraining the phase to that of the trial function as in the
fixed phase approximation \cite{ortiz93}. Both give about the same results,
within error bars, and we report values where the real part is positive.
For spin independent time reversal invariant potentials both reduce to the
fixed-node approximation. It is straightforward to show that if the
sign of the real part is that of the  correct ground state, we
will get the correct answer and small deviations give second
order corrections to the energy. We have not been able to prove that
this constraint always gives an upper bound to the ground state energy
although it appears to do so for the calculations we have done
to date. It seems likely that there is not an upper bound theorem for
the mixed estimate of the energy. If forward walking
or a path integral ground state technique
\cite{baroni99,pigs} is used, the method simply produces a better
trial function and the energy must be an upper bound.

In the fixed node method \cite{schmidt84}
the nodal structure of the trial function is determined by the Slater
Determinant. Similary, our path constraint is fully determined by the space
spin Slater Determinant of Eq. (\ref{jastrow}). The Jastrow function therefore
affects only the variance and not our final results.

\subsection{Tail corrections}

Monte Carlo calculations are generally 
performed within the sphere of radius $L/2$, where $L$ is the length of
the box side. Usually tail corrections are estimated 
by integrating out the spin--independent part of the 
two--body potential from $L/2$ up to infinity. We have made our
calculations within the full simulation box, and, in order to  
include also the contribution from the neighbor cells, 
we have tabulated the Jastrow factor $f(r)$ and 
the components $v_p(r)$ of the two-body potential in the following form
 
\begin{eqnarray}
F(x,y,z) &=& \prod_{mno} f
(|(x+mL_x)\hat x \nonumber \\
&+& (y+nL_y)\hat y +(z+oL_z)\hat z |) \nonumber\\
V_p(x,y,z) &=& \sum_{mno} v_p
(|(x+mL_x)\hat x \nonumber \\
&+& (y+nL_y)\hat y+(z+oL_z)\hat z |) \,.
\label{tail}
\end{eqnarray}

For the calculations shown, we found it adequate to include only the 26
additional
neighbor cells corresponding to $m$, $n$, and $o$ taking the values
$-1$, $0$, and $1$.

Our results are therefore already tail corrected. We found that 
the standard way of treating tail corrections leads to results very close to
ours, except when we consider model interactions which contain tensor forces,
which are relatively long range forces.

The three body potential is not treated as the two body one. 
Here we have estimated the tail corrections to the three body potential from
the PBFHC
variational results described in \ref{sec.fhnc}. This analysis shows
that such corrections are already very small for systems with 
66 nucleons.

\subsection{The Algorithm}

Finally, in this subsection we give the schematic structure 
of the AFDMC algorithm.

\begin{enumerate}

\item
Sample $|R,S \rangle$ initial walkers  from $|\langle \Psi_T | R, S \rangle|^2$
using Metropolis Monte Carlo.

\item
Propagate the spatial degrees of freedom
in the usual diffusion Monte Carlo way with a drifted
gaussian for half a time step.

\item
For each walker, diagonalize the potential matrix 
(two-- and three--body terms).

\item
Loop over the eigenvectors, sampling the corresponding Hubbard-Stratonovich
variable and update the spinors for half a time step. 
Introduce approximate importance sampling of the Hubbard-Stratonovich
variables, as discussed in the previous subsection.

\item Propagate the spin--orbit, using importance sampling.
\item
Repeat 2, 3, and 4 in the opposite order to produce a reversible propagator
to lower the time step error.

\item
Combine all weight factors and evaluate the new 
value of $\langle \Psi_T |  R, S \rangle$. If the real part is less than
0  include the walker in the evaluation of the mixed and the growth
energies, and then enforce constrained path by dropping the walker. 
In general, the
importance sampling makes the number of dropped walkers small.

\item
Evaluate the averages of 
$\langle \Psi_T |  R, S \rangle$, and 
$\langle \Psi_T|H|  R, S \rangle$ to calculate the mixed energy.
\item
Repeat as necessary.
\end{enumerate}

\section{FHNC and PBFHNC calculations}
\label{sec.fhnc}

In this Section we present the method that we have used to estimate 
the finite size effects in AFDMC simulations.
Such a method is made necessary by the fact that
simulations with more than 100 nucleons are computationally
very demanding.
A many--body theory, like FHNC, based on integral equation techniques, in which
the number of particles in the simulation box has no practical limitation seems
to be the best candidate to do this.
 
FHNC theory was originally developed
\cite{fantoni74} to treat 
fermionic systems in the thermodynamic limit.
However, FHNC has been recently reformulated to deal with a finite
number of fermions in a periodic box, as those used
for the Monte Carlo calculations \cite{fantoni01}.
Such a theory, denoted as Periodic Box FHNC (PBFHNC), is based upon
the fundamental property of the FHNC cluster expansion to be valid at 
all $1/A$ order
\cite{fantoni74,fantoni98}, and it has been developed
for Jastrow--correlated wave functions. In
the cases of a nucleonic system interacting via a central potential
it has been shown that finite size effects are (i) not limited to
the kinetic energy expectation value, and (ii) rather accurately
estimated by PBFHNC calculations \cite{manchester}.

However, realistic correlations $\hat{F}(ij)$ are spin--dependent and have an
operatorial structure similar to that of the two--body potential, as in 
Eq.(\ref{ham_arg}) (where the component $p=1$ corresponds to the Jastrow
correlation). Therefore the PBFHNC developed in Ref. \cite{fantoni01}
cannot be used as such, but has to be generalized to treat spin--dependent
correlations.
The main problem is that the spin operators involved do not commute,
namely $[\hat{F}(ij),\hat{F}(ik)] \neq 0 $. This feature
makes a full FHNC summation impossible and one has to resort to
reasonable approximations for the spin--dependent correlations.

Such approximations are characterized by the fact that,
whereas the cluster diagrams containing scalar correlations only are
summed up with FHNC technique, only a limited set of cluster diagrams 
containing spin--dependent correlations are included in the calculation.
The most tested and used approximation is the so called
FHNC/SOC, described in Ref. \cite{pandharipande79}. 
In our calculations we have used the version adopted in \cite{wiringa88}
in order to compute the different
correlation functions $\hat F(i,j)$ at the various densities considered. 
We have considered only the three variational parameters 
corresponding to healing distance, $d_c$, of central ($p=1-4$) and spin-orbit 
correlations ($p=7,8$), the healing distance, $d_t$ of tensor correlations 
($p=5,6$), and the quencher, $a_s$, of the spin-isospin dependent correlation.
The other variational parameters,
like the spin--independent potential quencher and the correlation quenchers
have been kept fixed at unity.
The {\em optimal} values of such variational parameters for pure neutron
matter are shown in Table \ref{tpar}. They have been obtained by minimizing the 
average energy $E_{av}=\frac{1}{2} (E_{JF}+E_{PB})$, where the two
energy expectation values $E_{JF}$ and $E_{PB}$ refer to the
Jackson--Feenberg and Pandharipande--Bethe kinetic energy 
expressions respectively \cite{pandharipande79}. The usual constraint 
$\frac{|E_{JF}-E_{PB}|}{E_{av}}\alt .005$ has been imposed in order to 
limit the range of variability of the free parameters 
in a region of reliability of the FHNC/SOC approximation.  
We have verified that in
such region the normalization condition is fulfilled within a few percent.

\begin{table}
\caption{\label{tpar}
Variational parameters used in our FHNC/SOC
and PBFHNC calculations for the $AU6'$ and $AU8'$ potentials.
$r_0=(3/(4\pi\rho))^{1/3}$ is the average distance between the 
neutrons. $r_0$, $d_c$ and $d_t$ are given in fm.
The reference density $\rho_0=0.16$ fm$^{-3}$ is the
equilibrium density of nuclear matter.
} 
\begin{ruledtabular}
\begin{tabular}{rrrrrrrr}
 $\rho/\rho_0$ & $r_0$ & $d_c(6)$ & $d_t(6)$ & $a_s(6)$ & 
$d_c(8)$ & $d_t(8)$ & $a_s(8)$  \\
\hline 
   $0.75$ & 1.258  & 1.761  & 4.695  & 0.9  & 2.264 & 4.528 & 0.8 \\    
   $1.00$ & 1.143  & 1.714  & 4.571  & 0.9  & 2.285 & 4.571 & 0.8 \\
   $1.25$ & 1.061  & 1.485  & 4.752  & 0.9  & 2.228 & 3.960 & 0.8 \\
   $2.0$  & 0.907  & 1.723  & 4.595  & 0.8  & 2.267 & 4.535 & 0.7 \\ 
   $2.5$  & 0.842  & 1.768  & 4.715  & 0.8  & 2.189 & 5.004 & 0.7 \\ 
\end{tabular}
\end{ruledtabular}
\end{table}

The variational energies for the case of the $AU6'$ interaction 
are reported on Table \ref{tfhnc6}. The Table also
reports the second order CBF perturbative corrections $\Delta E_2$ 
\cite{fabrocini02}
and the contribution from the lowest order elementary
diagram $\Delta E_{elem}$, as discussed in Ref. \cite{manchester}.
The not negligible value of $\Delta E_{elem}$ indicates that the
effect from elementary diagrams is larger than has been assumed
in all the past FHNC/SOC calculations of the nuclear matter equation
of state \cite{manchester,carlson03}.
In recent FHNC/SOC calculations of the equation of symmetric nuclear matter
and pure neutron matter \cite{akmal98,morales02} extra 
cluster diagrams with respect to the approximation used here have been 
included.
Differences between the various FHNC/SOC calculations are within the
predictive accuracy of the approximation.

\begin{table}
\caption{\label{tfhnc6}
FHNC/SOC energy per particle of neutron matter for the
$AU6'$ interaction, 
at various densities. $T_F$ is the Fermi
kinetic energy, and $\langle T \rangle$ is the kinetic energy expectation value,
corresponding to the average of the JF and PB kinetic energies.
$\langle V_2 \rangle$ and $\langle V_3\rangle$ 
are the expectation values of the 
two--body and three--body potentials respectively.
$\Delta E_2$ is the second order perturbative correction \cite{fabrocini02}.
$\Delta E_{elem}$ is the contribution from the lowest order
elementary diagram (see text). All the quantities, except $\rho/\rho_0$, 
are expressed in MeV.
} 
\begin{ruledtabular}
\begin{tabular}{rrrrrrrr}
 $\rho/\rho_0$ & $T_F$ & $\langle T \rangle $ & $\langle V_2\rangle $ 
               & $\langle V_3\rangle $ & 
$E_{\rm FHNC}$ & $\Delta E_2$ & $\Delta E_{elem}$  \\
\hline 
   $0.75$ & 28.969  & 35.33  & -22.67 & 2.58  & 15.2 & -0.9 & 0.6 \\    
   $1.00$ & 35.094  & 43.82  & -28.58 & 5.17  & 20.4 & -0.9 & 0.9 \\
   $1.25$ & 40.722  & 52.27  & -34.11 & 8.53  & 26.7 & -1.5 & 1.2 \\
   $2.0$  & 55.708  & 74.40  & -46.93 & 27.29 & 54.8 & -4.4 & 2.8 \\ 
   $2.5$  & 64.643  & 88.85  & -53.36 & 44.72 & 80.2 & -6.1 & 3.8 \\ 
\end{tabular}
\end{ruledtabular}
\end{table}

In Table \ref{tfhnc68} we compare the results of 
two different FHNC/SOC calculations of the
equation of state of neutron matter, carried out 
for the $AU8'$ potential. In the
first one ($AU8'(f_6)$) the spin--orbit correlation is set equal to zero,
whereas, in the second one ($AU8'(f_8)$), is included. 
One can see that the introduction of the spin-orbit correlation leads to
a large lowering in the energy.
As it will be shown, we do not find such a lowering when the spin-orbit
interaction is included in the AFDMC simulations. In the FHNC/SOC approximation
the cluster contributions from spin--orbit correlations
are correctly included only at the lowest order level. 
The many--body cluster contributions are essentially neglected.
The large and attractive 
spin--orbit contribution found in the $AU8'(f_8)$ calculation may be due
to this inaccuracy. On the other hand it might be possible that nodal surface
induced by the spin-orbit part of the interaction is not accurately described
by our trial function. 

\begin{table}
\caption{\label{tfhnc68}
Comparison of the FHNC/SOC results for the $AU8'$ interaction,
obtained with correlation operator of the type $f_6$ or of the
type $f_8$. In the first case the contribution of the spin--orbit potential
is calculated perturbatively from the $AU6'$ Hamiltonian. For comparison,
in the third column the results for the $AU6'$ interaction are also reported.
In all cases the contribution from elementary diagrams has been added.
} 
\begin{ruledtabular}
\begin{tabular}{rrrrr}
 $\rho/\rho_0$ & $T_F$ & $AU6'$ & $AU8'(f_6)$ & $AU8'(f_8)$  \\
\hline 
   $0.75$ & 28.969  & 15.8 & 16.1 & 13.3  \\    
   $1.00$ & 35.094  & 21.3 & 21.8 & 17.6  \\
   $1.25$ & 40.722  & 27.9 & 28.8 & 23.0  \\
   $2.0$  & 55.708  & 57.6 & 59.0 & 47.5  \\ 
   $2.5$  & 64.643  & 84.0 & 86.2 & 71.7  \\ 
\end{tabular}
\end{ruledtabular}
\end{table}

In order to compute the finite size effects in a realistic way
one should first generalize the PBFHNC theory to include SOC diagrams
like in FHNC/SOC approximation.  
Work in this direction is in progress \cite{arias03}.
In this paper we limit ourselves to including only
the two--body cluster diagrams for the
two--body potential and the kinetic energy and the leading three--body cluster
diagrams for the three--body potential \cite{arias03} in the PBFHNC scheme.
Such leading terms correspond to include up to two correlation operators
in the three-body cluster diagrams.
We will show that 
this approximation, hereafter denoted as PBFHNC/L, can already be used to 
roughly estimate the finite size effects.

\begin{table}
\caption{\label{tpbfhnc.16}
Comparison of the energy $E_2$ at the second order of the
FHNC cluster expansion with the full FHNC energy, $E_{\rm PBFHNC}$.
The calculation has been performed for the $v_6'$ model interaction at
$\rho=0.16$ fm$^{-3}$ and Jastrow correlation factor.
} 
\begin{ruledtabular}
\begin{tabular}{rrrr}
 $N$ & $T_F$ & $E_2$ & $E_{\rm PBFHNC}$ \\ 
\hline 
 $14$    & 35.600 & 19.36 & 17.60 \\
 $38$    & 33.703 & 17.51 & 15.91 \\
 $66$    & 34.917 & 19.11 & 17.63 \\
 $114$   & 35.646 & 20.09 & 18.71 \\
$1030$   & 35.139 & 19.46 & 18.04 \\
\end{tabular}
\end{ruledtabular}
\end{table}

\begin{table}
\caption{\label{tpb.16}
PBFHNC/L results for the $AU6'$ interaction at
density $\rho=0.16 $fm$^{-3}$. The Fermi kinetic energy $T_F$,
the expectation values of the kinetic energy $\langle T\rangle $, the two--body
potential $\langle V\rangle_2$ and the three--body potential 
$\langle V\rangle_3$ are 
displayed together with the energy per particle $E$ in MeV units. 
} 
\begin{ruledtabular}
\begin{tabular}{rrrrrr}
 $N$ & $T_F$ & $\langle T \rangle $ & $\langle V \rangle_2$ 
             & $\langle V\rangle_3$ & $E$  \\
\hline 
 $14$    & 35.600 & 44.47 & -29.41 & 4.31 &  19.37  \\
 $38$    & 33.703 & 42.41 & -29.43 & 4.70 &  17.68  \\
 $66$    & 34.917 & 43.64 & -29.07 & 4.82 &  19.39  \\
 $114$   & 35.646 & 44.40 & -28.87 & 4.87 &  20.40  \\
$1030$   & 35.139 & 43.88 & -28.95 & 4.86 &  19.79  \\
\end{tabular}
\end{ruledtabular}
\end{table}

The performance of the two-body cluster approximation to account for finite
size effects is studied in Table \ref{tpbfhnc.16}. There, for a purely
central potential without three-body force, PBFHNC/L and PBFHNC energies
are compared at $\rho=0.16$ fm$^{-3}$ for the range of particle numbers used
in our quantum Monte Carlo simulations.

Tables \ref{tpb.16} and \ref{tpb.32} give the PBFHNC/L results
for the $AU6'$ interaction
at two different densities for a number of neutron systems. 
Notice that the energy differences between the cases with 66 and 114
neutrons are very close to those obtained in the AFDMC simulations,  
given in Tables \ref{T_au6} and \ref{T_au8}. Systems with 14 and 38
neutrons are too small to be included in the finite size effects analysis.

\begin{table}
\caption{\label{tpb.32}
As in Table \ref{tpb.16} at density $\rho=0.32$ fm$^{-3}$. 
} 
\begin{ruledtabular}
\begin{tabular}{cccccc}
 $N$ & $T_F$ & $\langle T \rangle $ & $\langle V\rangle_2$ 
             & $\langle V \rangle_3$ & $\langle E\rangle$  \\
\hline 
 $14$    & 56.512 & 74.33 & -48.04 & 17.18 & 43.47  \\
 $38$    & 53.500 & 71.64 & -50.25 & 19.36 & 40.75  \\
 $66$    & 55.428 & 73.41 & -49.51 & 20.30 & 44.20  \\
 $114$   & 56.584 & 74.56 & -48.94 & 20.78 & 46.40  \\
$1030$   & 55.779 & 73.75 & -49.08 & 20.84 & 45.51  \\
\end{tabular}
\end{ruledtabular}
\end{table}

\section{Results}
\label{sec.results}
\subsection{AFDMC results for neutron matter}

Extensive neutron matter calculations have been carried out 
for the $AU6'$ and $AU8'$ interactions by considering 14, 38, 66 and 114
neutrons in a periodic box at various densities ranging from
$0.75\rho_0$ up to $2.5\rho_0$.

\begin{figure}[tb]
\includegraphics[width=\columnwidth,clip,bb= 110 447 478 671]{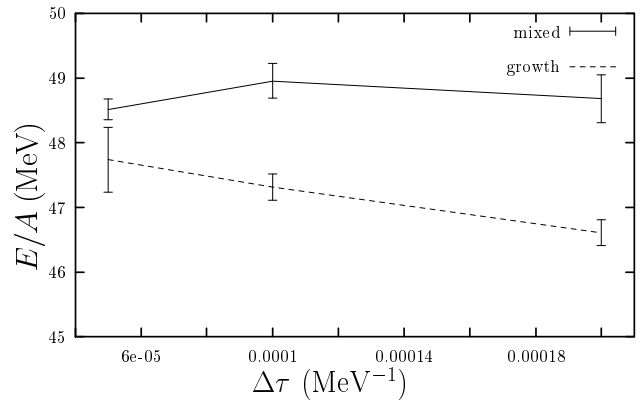}
\caption[]{
Mixed and Growth energies versus the time step for 14 neutrons at
$\rho=0.32$ fm$^{-3}$ with the $AU8'$ interaction. 
}
\label{fig1}
\end{figure}

In figure \ref{fig1} we show a typical behavior of the mixed and
growth energy as a function the time step for 14 neutrons in a periodic box
at $\rho=0.32$ fm$^{-1}$ interacting via $AU8'$.
At $\Delta \tau =5 \times 10^{-5}$  fm$^{-1}$
we have found that the statistical error are smaller than the extrapolation 
ones irrespective of the density and number of particles. 
All the calculations reported here have been obtained by using that value for
the time step.

The 14 neutrons system is interesting because it is small enough to be
studied by using other many body methods which become inefficient
for larger systems. In order to provide a full set of results for this
system in table \ref{tab.14neut} we report the energies at several densities 
calculated with the $v_6'$ and $v_8'$ interactions. 

\begin{table}
\caption{\label{tab.14neut}
AFDMC energies per particle in MeV of 14 neutrons in a periodic box for
interaction models at various densities. 
Error bars for the last digit are shown in parentheses.
}
\begin{ruledtabular}
\begin{tabular}{rrr}
$\rho$(fm$^{-3}$)  & $v_6'$ &  $v_8'$   \\
\hline
0.12 & 12.41(4) &  12.32(5)    \\
0.16 & 15.12(4) &  14.98(6)   \\
0.20 & 17.86(5) &  17.65(7)   \\
0.32 & 27.84(6) &  27.3(1)    \\
0.40 & 36.0(1)  &  35.3(1)   \\
\end{tabular}
\end{ruledtabular}
\end{table}

Diffusion Monte Carlo calculations using a pair-product wave function
for 14 neutron systems have just been reported \cite{carlson03}. They
however set the potential discontinuously to zero at distances
greater than $L/2$, while we use either the nearest image convention
or a lattice sum giving a continuous potential. We expect better
extrapolation to large system sizes
with the continuous potential as well as smaller time step errors.
The time step errors will affect our AFDMC calculations more because we
currently use the primitive approximation rather than
building the Green's
function from a product of exact two-body Green's functions. In principle
we could use the Hubbard-Stratonovich breakup for the pair-product
Green's function.
In any case, we have carried out a calculation at
$\rho=0.16$ fm$^-3$ using the same discontinuous potential and obtained
20.64(2) MeV and 20.32(6) MeV for the $v_6'$ and $v_8'$ potentials respectively
compared with their values of $19.91(11)$ and $17.00(27)$.
The larger difference when the spin-orbit term is included
in the Hamiltonian may be due to the different trial wave functions used.

In Tables \ref{T_au6} and \ref{T_au8} we report the results obtained 
with the AFDMC method of this work for neutron matter at the different 
densities considered for various system sizes.  The extrapolation to
infinite number of particles is carried out by using the PBFHNC/L results
for a given number of neutrons and for the infinite system. 

\begin{table*}
\caption{\label{T_au6}
AFDMC energies per particle in MeV for the $AU6'$
interaction obtained with systems with 14, 38, 66 and 114 neutrons at
various densities.
Error bars for the last digit of the Monte Carlo calculations 
are shown in parentheses.
The last column gives the extrapolated values from the PBFHNC/L 
calculation \cite{arias03}.
}
\begin{ruledtabular}
\begin{tabular}{ccccccc}
$\rho$(fm$^{-3}$) & AFDMC(14) & AFDMC(38) & AFDMC(66) & AFDMC(114)  
& AFDMC($\infty$)  \\
\hline
0.12            & 14.96(6)  & 13.76(9)  & 14.93(4) & 15.62(8)
& 15.0  \\
0.16            & 19.73(5)  & 18.56(8)  & 20.07(5) & 20.99(9)
& 20.4  \\
0.20            & 25.29(6)  & 24.4(1)   & 26.51(6) & 27.6(1)
& 26.9  \\
0.32            & 48.27(9)  & 49.8(1)   & 53.11(9) & 55.3(2)
& 54.4  \\
 0.40            & 69.9(1)   & 74.5(2)   & 79.4(2)  & 82.2(2)
& 81.3  \\
\end{tabular}
\end{ruledtabular}
\end{table*}

\begin{table*}
\caption{\label{T_au8}
AFDMC energies per particle in MeV for the $AU8'$
interaction obtained with systems with 14, 38 and 66 neutrons at
various densities. 
Error bars for the last digit of the Monte Carlo calculations 
are shown in parentheses.
The last column gives the extrapolated values from the PBFHNC/L 
calculation \cite{arias03}.
}
\begin{ruledtabular}
\begin{tabular}{ccccc}
$\rho$(fm$^{-3}$) & AFDMC(14) & AFDMC(38) & AFDMC(66)  
& AFDMC($\infty$) \\
\hline
0.12            & 14.80(9)  & 13.96(5)  & 15.26(5)  & 15.5 \\
0.16            & 19.76(6)  & 18.67(6)  & 20.23(9)  & 20.6 \\
0.20            & 25.23(8)  & 24.7(1)   & 27.1(1)   & 27.6 \\
0.32            & 48.4(1)   & 46.8(2)   & 54.4(6)   & 55.6 \\
0.40            & 70.3(2)   & 76.3(2)   & 81.4(3)   & 83.5 \\
\end{tabular}
\end{ruledtabular}
\end{table*}

\begin{figure}[tb]
\includegraphics[width=\columnwidth,clip,bb= 110 447 501 671]{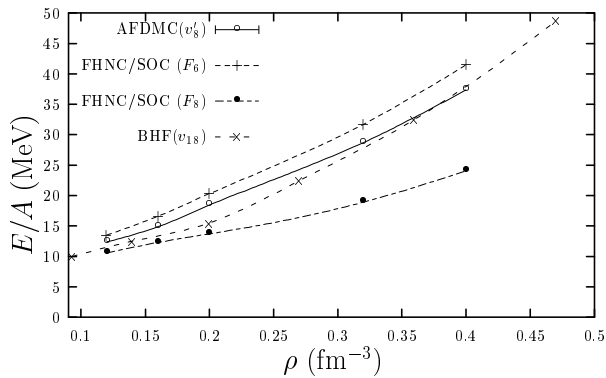}
\caption[]{
AFDMC energy per particle for neutron matter with obtained from 
simulations with 66 neutrons and the $v_8'$ potential. The 
variational FHNC/SOC results obtained with
correlation functions of type $F_6$ and $F_8$ are and the
Brueckner-Hartree-Fock (BHF) results of \cite{baldo00} also plotted and the
The statistical error in the AFDMC results are smaller than the symbols.
}
\label{fig2}
\end{figure}

The spin--orbit contribution is rather small at all of the densities considered.
This contrasts with previous FHNC/SOC calculations.
In Fig. \ref{fig2}, we plot the AFDMC results together with
the variational FHNC/SOC results for the $v_8'$ interaction
obtained by using correlation operators of the $F_6$ and $F_8$ forms,
and the Brueckner-Hartree-Fock (BHF) results for the $v_{18}$ potential
\cite{baldo00}. One can see that
SOC(F6) and SOC(F8) in the figure give quite different equations of state,
particularly at high density. We have tried transient estimates and AFDMC
simulations with orbitals of spin-backflow type for the 14 neutrons system
finding not more than roughly 5\% lowering of the energy \cite{brualla03}.

The three-body potential gives a large contribution to the
energy per particle at high densities.
Therefore the search for a realistic three--body
potential is a very fundamental problem for the study of dense and cold
hadronic matter. A considerable amount
of work has been done to find, in a semi--phenomenological way,  three--body
potentials to describe ordinary matter. However, whether such
potentials are also valid in the high density regime is still an open and 
debated issue. In table  \ref{illinois3b} we report AFDMC results 
performed with the two body $v_6'$ interaction and five different three-body
potentials including  the
Urbana IX \cite{pudliner95,pudliner97} (UIX) and the
recent Illinois 1 through 4 \cite{pieper01} three-body interactions.
One can see that already at
twice the nuclear matter density, the energy contributions from the
three--body potentials are large and very different from each other, in spite
of the fact that all of them provide a satisfactory fit to
the ground state and the low energy spectrum of nuclei with $A\le 8$.

\begin{table}[tb]
\caption{\label{illinois3b}
AFDMC energies per particle in MeV for the $v_6^{\prime}+$IL potentials
calculated with 66 particles. For the case of $v_6^{\prime}+$IL2 interaction,
at $\rho=0.32$ fm$^{-3}$ the energies per particle with 38 and 54 neutrons 
are 12.6(2) and 10.0(3) MeV respectively.
}
\begin{ruledtabular}
\begin{tabular}{rrrrrr}
$\rho$(fm$^{-3}$) & $AU6'$   & IL1    &  IL2     &  IL3     &  IL4    \\
\hline
0.16              & 20.07(5) & 11.2(1)& 11.39(8) & 12.0 (4) &10.5(2) \\
0.32              & 53.11(9) & 8.0(4) & 11.1(3)  & 14.7(3)  & 4.7(3)   \\
                  
\end{tabular}
\end{ruledtabular}
\end{table}

\begin{figure}[tb]
\includegraphics[width=\columnwidth,clip,bb= 110 447 501 671]{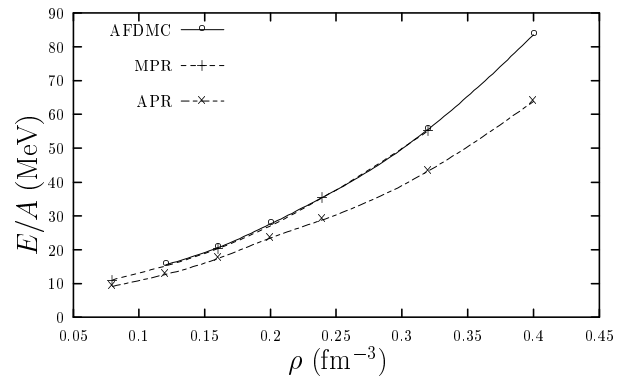}
\caption[]{
Extrapolated AFDMC Equation of state of pure neutron matter with the $AU8'$
potential (solid line). 
The variational results of Refs. \cite{akmal98} 
(APR, dotted-dashed line) and
\cite{morales02} (MPR, dashed line) corresponding to the 
Argonne $v_{18}$ -two body plus Urbana IX -three body potential are also 
plotted.
The lines are for guiding the eyes.
The statistical errors of the AFDMC estimates are smaller than the symbols.
}
\label{fig3}
\end{figure}

In Fig. \ref{fig3} we show the AFDMC equation of state for pure neutron 
matter with the $AU8'$ interaction corresponding to the extrapolated values
for infinite matter. We compare with the variational results of 
\textcite{akmal98} and the more recent ones of
Ref. \cite{morales02} both of them obtained with the Argonne 
$v_{18}$ 2- and Urbana IX 3- nucleon interactions.  One can see that there
is a surprising good agreement between our AFDMC results and the latest
variational calculation of \textcite{morales02}.

\begin{figure}[tb]
\includegraphics[width=\columnwidth,clip,bb= 103 447 501 671]{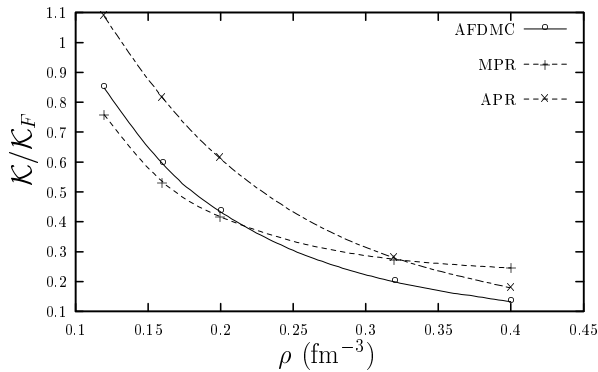}
\caption[]{
Compressibility ratio, ${\cal K}/{\cal K}_F$, for neutron matter 
obtained from the
extrapolated AFDMC energy per particle  with the $AU8'$
potential (solid line). 
The compressibility obtained from the variational results of 
Refs. \cite{akmal98} (APR, dotted-dashed line) and
\cite{morales02} (MPR, dashed line) is also plotted.
The lines are for guiding the eyes.
The statistical errors of the AFDMC estimates are smaller than the symbols.
}
\label{fig4}
\end{figure}

The compressibility ${\cal K}$, given by
 
\begin{equation}
\frac{1}{\cal K}=\rho^3~ \frac{\partial^2 E_0(\rho)}{\partial \rho^2}+
2\rho^2 \frac{\partial E_0(\rho)}{\partial \rho} \ ,
\end{equation}
can be estimated from the equation of state by taking $E_0=E/A$.
For a Fermi gas the compressibility is
${\cal K}_F = 9 \pi^2 m /(k_f^5 \hbar^2)$.
The AFDMC results for ${\cal K}/{\cal K}_F$ obtained from the extrapolated
energies with $AU8'$ 
are shown in Fig. \ref{fig4}.
They are compared with the compressibility calculated from the variational
energies of Ref. \cite{akmal98,morales02}.

\subsection{AFDMC results for nuclear matter}
 
The AFDMC can deal with $N\neq Z$ systems, and we have applied it to
compute the asymmetry coefficient of the mass formula for the semirealistic
two--body potential MS3 which is spin-isospin dependent but
has no tensor force \cite{afta68,guardiola81}
The resulting values of $E/A$ at $\rho_0$ for symmetrical nuclear matter
are given in Table \ref{tpb.MS3}, where they are also
compared with the FHNC/SOC
and PBFHNC results. The finite size correction is
estimated from the corresponding PBFHNC results.

\begin{table}[tb]
\caption{\label{tpb.MS3}
Finite size corrections for symmetrical nuclear matter \cite{manchester}:
PBFHNC results for the MS3 potential at $\rho=0.16$ fm$^{-3}$.
The PBFHNC calculations have been performed with a Jastrow correlated wave
function, whereas the FHNC/SOC result has been obtained with
a correlation operator of the type $F_4$. PBFHNC and FHNC/SOC calculations
include the basic four--point elementary diagram $E_4$.
}
\begin{ruledtabular}
\begin{tabular}{rrrr}
$ A $ & PB-FHNC  & FHNC/SOC   & AFDMC    \\
\hline
     28  & -13.6       &  -           &  -16.17(6) \\
     76  & -15.6       &  -           &  -18.08(3)    \\
   2060  & -14.0       &  -           &  $\downarrow$    \\
$\infty$ & -14.0       & -14.9        &  -16.5(1)    \\
\end{tabular}
\end{ruledtabular}
\end{table}
 
\begin{figure}[tb]
\includegraphics[width=\columnwidth,clip,bb= 103 452 496 671]{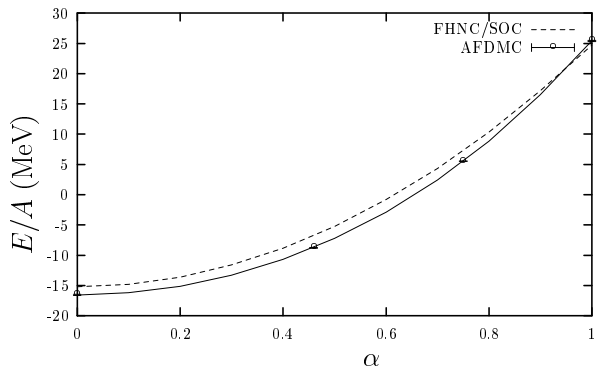}
\caption[]{
AFDMC and FHNC/SOC energy per particle of nuclear matter for several
values of the asymmetry parameter \cite{manchester}.
The lines correspond to polynomial fits of the calculated energies.
}
\label{fig5}
\end{figure}

In Fig. \ref{fig5} we plot the AFDMC energy per particle as a function 
of the asymmetry parameter, $\alpha =(N-Z)/(N+Z)$, of nuclear matter.
The FHNC/SOC curve corresponds to a quadratic fit of nuclear matter 
($\alpha=0$) and pure neutron matter ($\alpha=1$).

FHNC/SOC can only be used to study  $N=Z$ or $N=A$ matter. The symmetry
energy obtained from FHNC/SOC is $41.59$ MeV.
The function $E/A(\alpha)$ provided by the AFDMC results
is not fully quadratic in $\alpha$, and corresponds to
a symmetry energy of $\sim 36.4$  MeV.

\section{Outlook and Conclusions}
\label{sec.conclusions}

We have described a quantum Monte Carlo method specially suited to perform
calculations on nucleon systems with noncentral interactions. 
It has been applied here to calculate the equation of state of pure
neutron matter with fully realistic interactions by approximating it with
up to 114 neutron in a simulation box. Finite size effects have been 
estimated by performing 2- plus 3- body cluster diagrams calculations based
on PBFHNC method with spin-dependent correlations. The results obtained
show an overall agreement with Brueckner-Hartree-Fock calculations and
with a recent 2- plus 3- body cluster diagrams variational 
calculation \cite{morales02}. They also indicate that there is a very
small contribution coming from the spin-orbit component of the two-body
interaction while the effect from the three-body potential is quite large,
particularly at high densities. The large differences obtained for the
equation of state for different phenomenological three-body potentials
point out a three-body potential problem in the study of dense and cold
hadronic matter.

Work in progress is to validate the present results using trial wave
functions, other than the simple Slater determinant given in
Eq. (\ref{jastrow}),
that also can be calculated efficiently \cite{brualla03}. This will
allow us to both lower the variance of our calculations, as is usual when
better guiding functions are used in the importance sampling of the
random walk, as well as to obtain a better path constraint.

We believe that our method should be able to produce accurate Monte Carlo
calculations of a wide variety of nuclear systems. While previous Monte
Carlo calculations have been severely restricted on the particle number
by the spin-isospin sum this restriction is lifted by using the
auxiliary field breakup of the spin-isospin part of the Hamiltonian,
while using standard diffusion Monte Carlo for the spatial degrees of
freedom. A pressing need is simulating nuclear matter with fully
realistic interactions as already done for pure neutron matter;
calculating the properties
of light nuclei to compare with exact GFMC calculations; and investigating
pion condensation. In addition, including explicit meson degrees of freedom
can also be attempted. In the language of this paper,
each meson field mode corresponds to an auxiliary field \cite{schmidt99}.

\begin{acknowledgments}

We wish to thank A. Fabrocini, V. R. Pandharipande, A. Polls, 
S. Pieper, and R. Wiringa
for helpful conversations. 
S. F. wish to thank the International Centre for Theoretical Physics in 
Trieste for partial support. 
A. S. acknowledges the Spanish Ministerio 
de Ciencia y Tecnolog\'{\i}a for partial support under contract BMF2002-00200

\end{acknowledgments}

\appendix
\label{app.potpar}
\section{Potential Parameters}

The $A_{p,m}$ parameters in the expansion of the Argonne potentials,
Eq. \ref{ham_pot}, are shown in Table \ref{T_const}.
\begin{table*}[ht]
\caption{\label{T_const}
Argonne $v8'$ two--body potential. Matrix $A_{p,m}$
appearing in Eq.(\ref{ham_pot}).  
}
\begin{ruledtabular}
\begin{tabular}{rrrrrrrrr}
 p & $A_{p,1}$ & $A_{p,2}$ & $A_{p,3}$ & $A_{p,4}$ & $A_{p,5}$ &
$A_{p,6}$ & $A_{p,7}$ & $A_{p,8}$ \\ 
\hline
1 &    -7.52251741 &  2616.39024949 &     0 &   147.79390526 &     0 
& 0 &     0 &     0\\
2 &    -0.12318501 &    84.20118403 &     0 &   -61.22868919 &     0 
& 0 &     0 &     0\\
3 &     0.48726001 &   -82.48240972 &     0 &    49.26463509 &     0 
& 0 &     0 &     0\\
4 &     0.65399916 &  -107.98800762 &     0 &   -20.40956306 &     1/3 
&     2/3 &     0 &     0\\
5 &     0.94963459 &    -2.91931242 &  -424.28015518 &  -398.23289299 
&     0 &     0 &     0 &     0\\
6 &    -0.17865545 &    -0.97310414 &   234.18526077 &  -256.12175941 
&     0 &     0 &     1/3 &     2/3\\
7 &    -0.71193373 &  -373.43774331 &     0 &   653.08534247 & 0 &  0 
& 0 &     0\\
8 &    -0.28568125 &  -201.79028547 &     0 &   354.25604242 & 0 &  0 
& 0 &   0\\
\end{tabular}
\end{ruledtabular}
\end{table*}

The $F_m$ functions in Eq. \ref{ham_pot}, are written as follows

\begin{eqnarray}
F_1(r) & = & T^2(\mu,c;r) \ ,  \nonumber \\
F_2(r) & = & (1+a_0 r)W(r) \ , \nonumber \\ 
F_3(r) & = & \mu r W(r)    \ ,  \nonumber \\
F_4(r) & = & (\mu r)^2 W(r) \ , \nonumber \\ 
F_5(r) & = & a_1 Y(m_0,c;r) - a_2 r W(r)  \ ,  \nonumber \\
F_6(r) & = & a_3 Y(m_c,c;r) - a_4 r W(r)  \ ,  \nonumber \\
F_7(r) & = & a_1 T(m_0,c;r) \ , \nonumber \\ 
F_8(r) & = & a_3 T(m_c,c;r) \ , 
\label{ham_fun} 
\end{eqnarray} 
where the tensor, Yukawa and Wood--Saxon functions are defined as
\begin{eqnarray}
T(m,c;r) & = & (1+\frac{3}{mr}+\frac{3}{(mr)^2}) \frac{e^{-mr}}{mr}
(1-e^{-cr^2})^2 \ , \nonumber \\
Y(m,c;r) & = & \frac{e^{-mr}}{mr}(1-e^{-cr^2}) \ ,\nonumber \\
W(r) & = & \frac{1}{1+\exp(\ 5(r-0.5))}\ .
\label{yukawa}
\end{eqnarray} 

The the coefficients $a_0, \ldots, a_4$ are shown in Table \ref{acoef} and
the masses $m_0$, $m_c$ and $\mu$ and the cut--off parameter $c$ are given
in Table \ref{tmass}. 

\begin{table}
\caption{\label{acoef}
Argonne $v8'$ two--body potential. 
Values of the strength parameters in eqs. (\ref{ham_fun}) and(\ref{yukawa})
}
\begin{ruledtabular}
\begin{tabular}{rrrrr}
$a_0$ (fm$^{-1}$)  & $a_1$ & $a_2$(fm$^{-1}$) & $a_3$ & $a_4$ (fm$^{-1}$) \\
\hline
0.37929090 &     3.15588245 &    10.48427302 &     3.48918764 &    11.21004425\\
\end{tabular}
\end{ruledtabular}
\end{table}
\begin{table}
\caption{\label{tmass}
Argonne $v8'$ two--body potential. 
Values of the masses and cut-off parameters 
in eqs. (\ref{ham_fun}) and(\ref{yukawa})
}
\begin{ruledtabular}
\begin{tabular}{rrrr}
$m_0$ (fm$^{-1}$)  & $m_c$ (fm$^{-1}$) & $\mu$(fm$^{-1}$) & $c$(fm$^{-2}$)   \\
\hline
     0.68401113 &     0.70729025 &     0.69953054 &     2.1\\
\end{tabular}
\end{ruledtabular}
\end{table}

The values of the parameters of the three-body Urbana IX potential are 
shown in table \ref{tuix}

\begin{table}
\caption[]{\label{tuix}
Urbana IX three--body potential.
Values of the parameters appearing in Eq.(\ref{ham_three}).
}
\begin{ruledtabular}
\begin{tabular}{rrrr}
 $B_{2\pi}$ (MeV)  & $U_0$ (MeV)   & $m_{\pi}$(fm$^{-1}$)  
& $c_3$ (fm$^{-2}$)   \\
\hline
  -0.0586 & 0.0048 & 0.69953054 & 2.1  \\
\end{tabular}
\end{ruledtabular}
\end{table}

\end{document}